\documentclass[aps,prd,preprint,showpacs,showkeys]{revtex4}
\usepackage{bm}
\usepackage[mathscr]{euscript}
\begin{document}

\title{Berry phase of primordial scalar and
tensor perturbations in single-field inflationary models}

\author{Hamideh Balajany, Mohammad Mehrafarin}
\email{mehrafar@aut.ac.ir} 
\affiliation{Physics Department, Amirkabir University of Technology, Tehran
15914, Iran}

\begin{abstract}
In the framework of the single-field slow-roll inflation, we derive the Hamiltonian of the linear primordial scalar and tensor perturbations in the form of time-dependent harmonic oscillator Hamiltonians. We find the invariant operators of the resulting Hamiltonians and use their eigenstates to calculate the adiabatic Berry phase for sub-horizon modes in terms of the Lewis-Riesenfeld phase.  We conclude by discussing the discrepancy in the results of Pal et. al  [Class. Quant. Grav. {\bf 30}, 12 (2013)] for these Berry phases, which is resolved to yield agreement with our results.
\end{abstract} 


\maketitle
 
\section{Introduction}
Berry phase \cite{Berry} is a non-trivial geometric phase, distinct from the dynamical phase, that is picked up by a quantum system when it slowly traverses a closed path in the Hamiltonian parameter space.  Because of the wide range of its applications, examples of Berry phase have appeared in many different areas of physics and astronomy \cite{Shapere,Bohm,Mehrafarin1,Mehrafarin2,Torabi1,Torabi2,Bakke,Cai,Cai:1990,Corichi,
Mazur,Dutta,Melo,Melo2,Bakke2}. Of particular relevance to our work is the Berry phase of primordial cosmological perturbations, which are well accomodated in inflationary models \cite{Guth,Bassett,Mukhanov}. In single-field inflation, using the gauge invariant variable of Bardeen \cite{Bardeen:1980}, the Berry phase has been obtained  from the wave function of the perturbations by solving the associated Shr\"{o}dinger equation \cite{Pal}.  
As the origin of our present universe, primordial perturbations  have presumably left their mark  to  be traced in cosmological observations. In this regard, the Berry phase, as a footprint of the perturbations, can serve  to probe the cosmological inflation \cite{Campo}. 

In this work, we obtain the Berry phase of the linear primordial perturbations in the single-field slow-roll inflation via a different approach. Our approach is based on reducing the problem to a time-dependent harmonic oscillator and, thereby, using the Lewis-Riesenfeld invariant operator method \cite{Lewis:1968,Lewis,Carvalho,Pedrosa,Pedrosa2} to obtain the Berry phase. This approach  has been employed to obtain the Berry phase of relic gravitons in the FRW background \cite{Bakke}. Here, using the  gauge invariant variables of Malik and Wands \cite{Malik}, we derive the Hamiltonian of the scalar and tensor Fourier modes in the form of time-dependent harmonic oscillator Hamiltonians (Section 2). The Berry phase of a generalized harmonic oscillator has been derived in \cite{Monteoliva} using the Lewis-Riesenfeld invariant operator method. In the same manner, we find the invariant operators of the resulting Hamiltonians and use their eigenstates to calculate the adiabatic Berry phase for sub-horizon scalar and tensor modes as a Lewis-Riesenfeld phase (Section 3). Finally, we discuss the discrepancy in the results of \cite{Pal} for these Berry phases, which is resolved to yield agreement with our results.

\section{The perturbation Hamiltonian}

 In the single-field model, the universe is dominated by a scalar field ${\bar\varphi}$ with potential $V({\bar\varphi})$. The action is
\begin{equation}
S=\int  d^4x\sqrt{-g}\,\frac{1}{2}\left[R-g^{\mu\nu}\partial_\mu\bar\varphi\partial_\nu\bar\varphi-2V(\bar\varphi)\right]  \label{A}
\end{equation}  
where units have been chosen such that $8 \pi G=\hbar=c=1$. 
The background universe is the flat FRW spacetime
$$
ds^2=-{N}^2(t)dt^2+a^2(t)\delta_{ij}\label{metric}dx^idx^j
$$
where $a$ is the scale factor and $N$ depends on the choice of the time variable. (Conformal and cosmic time correspond to $N=a$ and $N=1$, respectively.) The background scalar field, which depends only on time, is  $\varphi(t)$ with conjugate momentum $\Pi=\dot{\varphi}/{N}$. In the ADM formalism \cite{Arnowitt}, where
$$
ds^2=-\bar {N}^2dt^2+\bar{h}_{ij}(dx^i+N^i dt)(dx^j+N^jdt)
$$
the perturbed universe has $\bar h_{ij}=a^2e^{2\alpha}\delta_{ij}+\gamma_{ij}$, where $\alpha(t,\bm{x})$ is the scalar curvature perturbation and $\gamma_{ij}(t,\bm{x})$ is a divergence-less and traceless metric perturbation that represents transverse gravity waves.

Let us first consider the scalar perturbations. The linear scalar gauge invariant perturbation variable is constructed from the curvature and field perturbations ($\alpha$ and $\delta\bar\varphi$) according to \cite{Malik}
$$
\zeta(t,{\bm x})=\alpha-\frac{H}{\Pi} \delta\bar\varphi 
$$
where $H(t)=\dot{a}/Na$ is the background Hubble parameter. 
The first order slow-roll parameters are given by
$$
\eta(t)=\frac{1}{NH} \frac{\dot{\Pi}}{\Pi },\ \ \
\epsilon (t) = -\frac{\dot{H}}{NH^2}.
$$
Working in the uniform energy density gauge, $\delta\bar\varphi=0$,  action (\ref{A}) to the second order in perturbation variable $\zeta$ is given by \cite{Tzavara}
\begin{equation}
S_{\text{scalar}}=\int  d^4x\, [a^{3}\frac{\epsilon}{N}{(\partial_t \zeta)}^2-a\epsilon N{(\partial_i\zeta)}^{2} ]. \label{E}
\end{equation}
Choosing $t$ to be the conformal time $\tau$ by setting $N=a$, and defining the Mukhanov-type variable  ${q}=-a\sqrt{2\epsilon}{\zeta}$, (\ref{E}) becomes
\begin{eqnarray} \begin{array}{c}
S_{\text{scalar}}=\int d\tau d^{3}x\, \frac{1}{2} [ {- {({\partial_i q})}^{2}
+{q^\prime}^ 2+{{\bar{\mathscr{H}}}^2}{q^2}-2{\bar{\mathscr{H}}}qq^{\prime}}],\\  \bar{\mathscr{H}}=\mathscr{H}+\frac{\epsilon^\prime}{2\epsilon}=\mathscr{H} (1+\epsilon+\eta)\label{ac}
\end{array}
\end{eqnarray}
where prime indicates conformal time derivative and $\mathscr{H}={a^{\prime}}/a=aH$ is the conformal Hubble parameter.  Representing the Fourier transforms of $q$ by $q_{\bm k}$ and forming the row matrix ${\bm q}_{\bm k}^T=(q_{\bm k}^{(R)}\ \ q_{\bm k}^{(I)})$ from the real and imaginary parts of $q_{\bm k}$, (\ref{ac}) can be written as 
$$
{S}_{\text{scalar}}=\int d\tau  \frac{d^3k}{(2\pi)^3}\mathcal {L}_{\bm{k},\text{scalar}}  \ \ , \ \mathcal {L}_{\bm{k},\text{scalar}}=\frac{1}{2}[({\bm q}_{{\bm k}}^{\prime}-\bar\mathscr{H}{\bm q}_{{\bm k}})^T({\bm q}_{{\bm k}}^{\prime}-\bar\mathscr{H}{\bm q}_{{\bm k}})-k^2{\bm q}_{{\bm k}}^{T}{\bm q}_{{\bm k}}].$$
The corresponding Hamiltonian is given by
$$
{\mathcal {H}}_{\text{scalar}}=\int  \frac{d^3k}{(2\pi)^3}{\mathcal {H}}_{\bm k,\text{scalar}}\ \ , \ 
{\mathcal {H}}_{\bm k,\text{scalar}}=\sum_m{{\bm p}}^T_{\bm k}{\bm q}^\prime_{\bm k}-{\mathcal{L}}_{\bm k}
$$
with ${\bm p}_{\bm k}^T=\partial {\mathcal{L}}_{\bm k,\text{scalar}}/\partial {\bm q}^{\prime}_{\bm k}=(p_{\bm k}^{(R)}\ \ p_{\bm k}^{(I)})$. 
Thus, promoting the canonically conjugate variables to operators (denoted by hat), the matrices become matrix operators, and
\begin{equation}
\hat{{\mathcal{\bm H}}}_{\bm k,\text{scalar}}=\frac{1}{2}[{\hat{\bm p}}^{T}_{\bm k}{{\hat{\bm p}}_{\bm k}}+\bar\mathscr{H}({{\hat{\bm p}}^{T}_{\bm k}}{{\hat{\bm q}}_{\bm k}}+{{\hat{\bm q}}^{T}_{\bm k}}{{\hat{\bm p}}_{\bm k}})+k^{2}{{\hat{\bm q}}^{T}_{\bm k}{\hat{\bm q}}_{\bm k}}]\label{H}
\end{equation}
which represents a time-dependent harmonic oscillator of frequency $\omega_{k}(\tau)=\sqrt{{k^{2}-{\bar\mathscr{H}}^{2}}}$.

As for the linear tensor perturbations, the second order action calculated from (\ref{A}) is \cite{Tzavara}
$$
{S}_{\text{tensor}}= \int {d^{4}x}\, \frac{1}{2}\,[  \frac{a^{3}}{4N}(\partial_t\gamma_{ij})^{2}-\frac{a N}{4}{({\partial_k\gamma}_{ij})}^{2}].
$$
Set $N=a$ and write the Fourier transforms $\gamma_{ij\bm k}$ in terms of the polarization tensors $\varepsilon_{ij}^{s}({\bm k})$ ($s=1,2$) as $\gamma_{ij\bm k}=\sum_s\frac{\surd 2}{a}\chi^s_{\bm k}\, \varepsilon_{ij}^{s}({\bm k})$. We similarly get
$$
{S}_{\text{tensor}}=\int d\tau  \frac{d^3k}{(2\pi)^3}\mathcal {L}_{\bm{k},\text{tensor}} \ \ , \ \mathcal {L}_{\bm{k},\text{tensor}}=\sum_{s=1}^2\frac{1}{2}[({\bm \chi}^{s\prime}_{\bm k}-\mathscr{H}{\bm \chi}^s_{\bm k})^{T}({\bm \chi}^{s\prime}_{\bm k}-\mathscr{H}{\bm \chi}^s_{\bm k})-k^2{\bm \chi}^{sT}_{\bm k}{\bm \chi}^s_{\bm k}]
$$
where $\bm {\chi}_{\bm k}^{sT}=(\chi_{\bm{k}}^{s(R)}\ \ \chi_{\bm{k}}^{s(I)})$. Note that the summation over $s$ pertains only when both polarizatios are present in the gravitational wave. Hence, defining the conjugate momenta $\bm{\pi}_{\bm{k}}^{sT}=\partial {\mathcal{L}}_{\bm k, \text{tensor}}/\partial {\bm \chi}^{s\prime }_{\bm k}=(\pi_{\bm{k}}^{s(R)}\ \ \pi_{\bm{k}}^{s(I)})$ and promoting to operators, we find
\begin{equation}
\hat{{\mathcal{\bm H}}}_{\bm k,\text{tensor}}=\sum_{s}\hat{{\mathcal{\bm H}}}^s_{\bm k,\text{tensor}}\ \ , \
\hat{{\mathcal{\bm H}}}^s_{\bm k,\text{tensor}}=\frac{1}{2}[ \hat{\bm \pi}^{sT}_{\bm k}\hat{\bm \pi}^s_{\bm k}+\mathscr{H}(\hat{\bm \pi}^{sT}_{\bm k}\hat{\bm \chi}^s_{\bm k}+\hat{\bm \chi}^{sT}_{\bm k}\hat{\bm \pi}^s_{\bm k})+k^{2}\hat{\bm \chi}^{sT}_{\bm k}\hat{\bm \chi}_{\bm k}^s]. \label{I}
\end{equation}
Thus, the Hamiltonian for tensor modes also coincides with that of a harmonic oscillator of frequency $\Omega_{k}(\tau)=\sqrt{{k^{2}-{\mathscr{H}}^{2}}}$.

\section{Berry phase of the scalar and tensor modes}

We use the invariant operator method \cite{Lewis:1968,Lewis} to determine the dynamical invariants of the harmonic oscillator Hamiltonians (\ref{H}) and (\ref{I}). The Berry phase can then be obtained as a Lewis-Riesenfeld phase \cite{Monteoliva}, which is constructed from the eigenstates of the invariant operator. 

The invariant operator, by definition, satisfies the von Neumann equation. It has been derived for the generalized harmonic oscillator Hamiltonian in the form,
$
\frac{1}{2}{[Z\hat{\bm p}^2+Y({{\hat{\bm p}}{\hat{\bm q}}}+{{\hat{\bm q}}{\hat{\bm p}}})+X\hat{\bm q}^2]},
$
where $X,Y,Z$ are time dependent \cite{Engineer}. This has the same form as Hamiltonians (\ref{H}) and (\ref{I}). Thence, for (\ref{H}) the invariant takes the form
$$
\hat{I}_{\bm k,\text{scalar}}=\frac{1}{2}\bigg\lbrace{ \frac{1}{\rho_{k}^2}\hat{\bm q}_{\bm k}^T\hat{\bm q}_{\bm k}+[{\rho_{k}(\hat{\bm p}_{\bm k}+\bar\mathscr{H}\hat{\bm q}_{\bm k}) - \rho_{k}^\prime \hat{\bm q}_{\bm k}}]^T[{\rho_{k}(\hat{\bm p}_{\bm k}+\bar\mathscr{H}\hat{\bm q}_{\bm k})- \rho_{k}^\prime \hat{\bm q}_{\bm k}}]}\bigg\rbrace
$$
where  the auxiliary variable ${\rho}_{k}(\tau)$ is a time-periodic solution of the Milne-Pinney equation
\begin{equation}
{{\rho}^{{\prime}{\prime}}_{k}}+({\omega}_{k}^{2}-{\bar\mathscr{H}}^\prime){{\rho}_{k}}-{\rho}^{-3}_{k}=0.\label{Milne}
\end{equation}

We define the raising and lowering matrix operators by
\begin{equation}
\hat{\bm A}_{\bm k}^{(\pm)}=\frac{1}{\sqrt{2}}\bigg\lbrace\frac{1}{\rho_{k}}\hat{\bm q}_{\bm k}\pm i[{  \rho_{k}^\prime \hat{\bm q}_{\bm k}}-\rho_{k}(\hat{\bm p}_{\bm k}+\bar\mathscr{H}\hat{\bm q}_{\bm k})]\bigg\rbrace \label{L}
\end{equation}
and write $\hat{\bm A}_{\bm k}^{(\pm)T}=(\hat{A}_{\bm{ k} 1}^{(\pm )}\ \ \hat{A}_{\bm{ k} 2}^{(\pm )})$. The components 1 and 2 are standard raising and lowering operators that satisfy
\begin{eqnarray}\begin{array}{c}
[\hat{A}_{\bm{k}1}^{(\pm )},\hat{A}^{(\pm )}_{\bm{k}2}]=0,   \ \ [\hat{A}_{\bm{k}1}^{(-)},\hat{A}^{(+)}_{\bm{k}1}]=[\hat{A}_{\bm{k}2}^{(-)},\hat{A}^{(+)}_{\bm{k}2}]=1 \\
\hat{A}_{\bm{k}1,2}^{(-)}\vert n_{\bm{k}1,2}\rangle =\sqrt{n_{\bm{k}1,2}}\, \vert n_{\bm{k}1,2}-1\rangle, \ \
\hat{A}^{(+)}_{\bm{k}1,2}\vert n_{\bm{k}1,2}\rangle =\sqrt{n_{\bm{k}1,2}+1}\, \vert n_{\bm{k}1,2}+1\rangle \label{N}
\end{array}
\end{eqnarray}
where $\vert{n_{{\bm k}1},n_{{\bm k}2}}\rangle$ is the eigenstate of $\hat{I}_{\bm k}^{\text{scalar}}=\hat{\bm A}_{\bm k}^{(+)T}\hat{\bm A}_{\bm k}^{(-)}+1$ with eigenvalue $n_{\bm{ k}1}+n_{\bm{ k}2}+1$.

The accumulated Berry phase over time period $\tau_0$  is derivable from the Lewis-Riesenfeld phase according to \cite{Monteoliva}
\begin{equation}
\Gamma_{\bm k,\text{scalar}}({n}_{\bm {k}1},{n}_{\bm {k}2},\tau_0)={\int}^{\tau_0}_{0}\left\langle {n}_{\bm{ k}1},{n}_{\bm {k}2} \left\vert i\partial_\tau\right\vert {n}_{\bm {k}1},{n}_{\bm{ k}2}\right\rangle\,  d\tau.
 \end{equation}
To calculate the integrand, we proceed as follows. From (\ref{N}), differentiation with respect to $\tau$ yields
$$
\frac{1}{\sqrt{n_{\bm {k}1}}}\left\langle{n_{\bm{ k}1}}\left\vert\partial_\tau \hat{A}^{(+)}_{\bm {k}1}\right\vert{n_{\bm {k}1}-1}\right\rangle=\left\langle{n_{\bm{ k}1}}\left\vert\partial_\tau\right\vert{n_{\bm{ k}1}}\right\rangle-\left\langle{n_{\bm {k}1}-1}\left\vert\partial_\tau\right\vert{n_{\bm{ k}1}-1}\right\rangle
$$
together with a similar expression with subscript $1$ replaced by $2$. It follows that
$$\left\langle{\bar n_{\bm {k}1}}\left\vert\partial_\tau\right\vert{\bar n_{\bm {k}1}}\right\rangle-\left\langle 0\left\vert\partial_\tau\right\vert 0\right\rangle=\sum_{n_{\bm{ k}1}=1}^{\bar n_{\bm {k}1}}
\frac{1}{\sqrt{n_{\bm{ k}1}}}\left\langle{n_{\bm {k}1}}\left\vert\partial_\tau \hat{A}^{(+)}_{\bm{ k}1}\right\vert{n_{\bm {k}1}-1}\right\rangle.
$$
By using (\ref{L}), we can express $\partial_\tau \hat{A}^{(+)}_{\bm {k}1}$ in terms of the raising and lowering operators to find
$$
\left\langle{n_{\bm {k}1}}\left\vert\partial_\tau \hat{A}^{(+)}_{\bm{ k}1}\right\vert{n_{\bm {k}1}-1}\right\rangle =-\frac{i}{2}\,(\omega^2_k\rho_{k}^2-\rho_{k}^{-2}+
{\rho_{k}^\prime}^2)\sqrt{n_{\bm {k}1}}
$$
and therefore
$$
\left\langle{n_{\bm {k}1}}\left\vert i\partial_\tau\right\vert{n_{\bm {k}1}}\right\rangle=\left\langle 0\left\vert i\partial_\tau\right\vert 0\right\rangle
+\frac{1}{2}\,(\omega^2_k\rho_{k}^2-\rho_{k}^{-2}+
{\rho_{k}^\prime}^2)\,n_{\bm{ k}1}.
$$
Bearing in mind the same expression with subscript $1$ replaced by $2$, it follows that
$$
\left\langle{n}_{\bm {k}1}, n_{\bm{k}2}\left\vert i\partial_\tau\right\vert , n_{\bm{k}1}, n_{\bm{k}2}\right\rangle=2\left\langle 0\left\vert i\partial_\tau\right\vert 0\right\rangle
+\frac{1}{2}\,(\omega^2_k\rho_{k}^2-\rho_{k}^{-2}+
{\rho_{k}^\prime}^2)\,(n_{\bm{ k}1}+n_{\bm {k}2}).
$$
Conveniently choosing the Lewis gauge \cite{Lewis} 
$$
\left\langle{0}\left\vert{i\partial_\tau}\right\vert{0}\right\rangle
= \frac{1}{4}\, (\omega^2_k\rho_{k}^2-\rho_{k}^{-2}+
{\rho_{k}^\prime}^2)
$$
we finally obtain 
\begin{equation}
{\Gamma}_{\bm k,\text{scalar}}=\frac{1}{2}(n_{\bm{k}1}+n_{\bm{k}2}+1)\int_0^{\tau_0} (\omega^2_k\rho_{k}^2-\rho_{k}^{-2}+
{\rho_{k}^\prime}^2)\,d\tau. \label{phase}
\end{equation}

In the adiabatic limit of slow time variation, we introduce the adiabatic parameter $\lambda$ ($\ll1$) and write $\eta=\lambda\tau$. Substituting in (\ref{Milne}) gives ${\rho}_{k}^2=1/\sqrt{k^2-\mathscr{H}^2}+O(\lambda)$ as $\bar\mathscr{H}^\prime=\lambda d\mathscr{H}/d\eta+O(\lambda^2)$, $\epsilon$ and $\eta$ being  first order in $\lambda$. Thus, on using (\ref{Milne}), the integrand of (\ref{phase}) becomes
$$
\bar\mathscr{H}^\prime\rho_k^2+{\rho_k^\prime}^2-\rho_k\rho_k^{\prime\prime}= \lambda \frac{d\mathscr{H}/d\eta}{\sqrt{k^2-\mathscr{H}^2}}+O(\lambda^2)\rightarrow\frac{\mathscr{H}^\prime}{\sqrt{k^2-\mathscr{H}^2}}.
$$
Hence, in the adiabatic limit,
$$
{\Gamma}_{\bm k,\text{scalar}}=\frac{1}{2}(n_{\bm{k}1}+n_{\bm{k}2}+1)\sin^{-1} \frac{\mathscr{H}_0}{k}
$$
where $\mathscr{H}_0=\mathscr{H}(\tau_0)$.
Note that ${k}\geq\mathscr{H}$, which means that the above result for Berry phase holds for sub-horizon modes that oscillate with real frequency. Thus, ${{\tau}_{0}}$ corresponds to the conformal time at which $\mathscr{H}=k$, i.e., $\mathscr{H}_0=k$, so that
\begin{equation}
{\Gamma}_{\bm k,\text{scalar}}= (n_{\bm{k}1}+n_{\bm{k}2}+1)\, \frac{\pi}{4}
\end{equation}
which yields $\Gamma_{\text{scalar}}=\pi/4$ for the ground state. The adiabatic Berry phase is, thus, independent of the (conformal) Hubble parameter, in contrast to the general non-adiabatic Berry phase given by (\ref{phase}). 

For tensor modes, because of the identical form of the Hamiltonian for each polarization state, as given by (\ref{I}) , we just have to introduce the polarization  index $s$ in the above steps and make the correspondences $\omega_k\rightarrow\Omega_k$, $\bar\mathscr{H}\rightarrow\mathscr{H}$. Thus the invariant operator of $\hat{{\mathcal{\bm H}}}_{\bm k,\text{tensor}}$ is $\hat{I}_{\bm k,\text{tensor}}=\sum_s \hat{I}_{\bm k,\text{tensor}}^{s}$, where $\hat{I}_{\bm k,\text{tensor}}^{s}$ is the invariant of $\hat{{\mathcal{\bm H}}}^{s}_{\bm k,\text{tensor}}$. We have  
$$
\hat{I}_{\bm k,\text{tensor}}^{s}=\hat{\bm A}_{\bm k}^{s(+)T}\hat{\bm A}_{\bm k}^{s(-)}+1, \ \ \ \hat{\bm A}_{\bm k}^{s(\pm)}=\frac{1}{\sqrt{2}}\bigg\lbrace\frac{1}{\rho_{k}}\hat{\bm \chi}^s_{\bm k}\pm i[{  \rho_{k}^\prime \hat{\bm \chi}^s_{\bm k}}-\rho_{k}(\hat{\bm \pi}^s_{\bm k}+\mathscr{H}\hat{\bm \chi}^s_{\bm k})]\bigg\rbrace
$$
where $\rho_k$ satisfies 
$$
{{\rho}^{{\prime}{\prime}}_{k}}+({\Omega}_{k}^{2}-{\mathscr{H}}^\prime){{\rho}_{k}}-{\rho}^{-3}_{k}=0.
$$
The eigenstate of $ \hat{I}_{\bm k,\text{tensor}}^{s}$ is $\vert\bm{n}^s_{\bm {k}}\rangle\equiv\vert n^s_{\bm{k}1},n^s_{\bm{k}2}\rangle$ with eigenvalue $ n^s_{\bm{k}1}+n^s_{\bm{k}2}+1$, so that the eigenstate of $\hat{I}_{\bm k,\text{tensor}}$  is $\vert\bm{n}_{\bm {k}}^1,\bm{n}^2_{\bm {k}}\rangle$. The Berry phase is, therefore, given by
$$
\Gamma_{\bm k,\text{tensor}}(\bm{n}^1_{\bm {k}},\bm{n}^2_{\bm {k}},\tau_0)={\int}^{\tau_0}_{0}\left\langle \bm{n}^1_{\bm{ k}},\bm{n}^2_{\bm {k}} \left\vert i\partial_\tau\right\vert\bm {n}^1_{\bm {k}},\bm{n}^2_{\bm{ k}}\right\rangle\,  d\tau.
$$
Noting that the integrand is equal to $\sum_s\left\langle{\bm{n}^s_{\bm k}}\left\vert i\partial_\tau\right\vert{\bm{n}^s_{\bm k}}\right\rangle$, we similarly obtain in place of (\ref{phase}),
$$
{\Gamma}_{\bm k,\text{tensor}}=\frac{1}{2}\sum_{s=1}^2(n_{\bm{k}1}^{s}+n_{\bm{k}2}^{s}+1)\int_0^{\tau_0} (\omega^2_k\rho_{k}^2-\rho_{k}^{-2}+
{\rho_{k}^\prime}^2)\,d\tau
$$
and hence, in the adiabatic limit,
\begin{equation}
{\Gamma}_{\bm k,\text{tensor}}=\sum_s(n_{\bm{k}1}^{s}+n_{\bm{k}2}^{s}+1)\,\frac{\pi}{4}.
\end{equation} 
The summation pertains only when both polarizations are present in the gravitational wave. For the ground state, therefore, we have  $\Gamma_{\text{tensor}}=\pi/4$ for each polarization.

\section{discussion}
Considering linear primordial perturbations in the single-field slow-roll inflation, we have derived the Hamiltonian of the scalar and tensor modes  in the form of time-dependent harmonic oscillator Hamiltonians. We obtained the invariant operators of the resulting Hamiltonians and used their eigenstates to calculate the adiabatic Berry phase for sub-horizon  perturbations as a Lewis-Riesenfeld phase. 

In conclusion, we ought to comment on the discrepancy in the results of \cite{Pal}, where the scalar and tensor adiabatic Berry phases are obtained from the wave function of the perturbations. Their results for the ground state  read as follows (in our notation): 
$$
\Gamma_{\text{scalar}}=-\frac{\pi}{4}\frac{1+3\epsilon-\eta}{\sqrt{1+2(3\epsilon-\eta})}+O(\epsilon^2,\eta^2,\epsilon\eta),\ \ \
\Gamma_{\text{tensor}}=-\frac{\pi}{4}\frac{1+\epsilon}{\sqrt{1+2\epsilon}}+O(\epsilon^2,\eta^2,\epsilon\eta)
$$
where $\Gamma_{\text{tensor}}$ pertains to each polarization.  They also relate the Berry phases to observable parameters, viz spectral indices, through the slow roll parameters $\epsilon, \eta$. In accordance with the adiabatic requirement, the above expressions are claimed by the authors to be exact to first order in $\epsilon, \eta$. This is obviously incorrect  because of the denominators. In fact, by a simple binomial expansion, the correct first order results are $\Gamma_{\text{scalar}}= \Gamma_{\text{tensor}}=-\pi/4$, which coincide with ours (up to an unimportant sign). Moreover, there is no relationship with spectral indices as far as the adiabatic approximation is concerned.

\end{document}